\documentclass[preprint,10pt,authoryear]{elsarticle}
\usepackage{graphicx, verbatim, url}

\journal{arXiv.org}

\begin{document}
\begin{frontmatter}

\title{Collaboration in computer science: \\ a network science approach. Part II}

\author{Massimo Franceschet}

\address{Department of Mathematics and Computer Science, University of Udine \\
           Via delle Scienze 206 -- 33100 Udine, Italy \\
           \texttt{massimo.franceschet@uniud.it}}

\begin{abstract}
We represent collaboration of authors in computer science papers in terms of both affiliation and collaboration networks and observe how these networks evolved over time since 1960. We investigate the temporal evolution of bibliometric properties, like size of the discipline, productivity of scholars, and collaboration level in papers, as well as of large-scale network properties, like reachability and average separation distance among scientists, distribution of the number of scholar collaborators, network clustering and network assortativity by number of collaborators.
\end{abstract}

\begin{keyword}
Bibliometrics; Research collaboration; Affiliation networks; Collaboration networks; Network science
\end{keyword}

\end{frontmatter}

\section{Introduction}

The process of scholarly publication of research achievements leaves behind itself a trail of footprints that can be weaved into densely intertwined bibliographic networks. Examples of such networks that have been extensively studied are \textit{citation networks}, which are webs of bibliographic references among academic papers or journals, and \textit{collaboration networks}, where scholars are connected if they co-authored a paper together. The field of \textit{network science} -- the holistic analysis of complex systems through the study of the structure of networks that wire their components -- is an optimal solution for the study of these (and many other) networks \citep{N10}.

We take advantage of the network science paraphernalia and make a longitudinal (time-resolved) study of properties that emerge from co-authorship in conference and journal publications in computer science. We represent co-authorship in terms of two differently grained networks. The first representation is an author-paper \textit{affiliation network}, that is a bipartite graph with two series of nodes representing authors and papers, and links running from authors to papers they authored. With the support of this network we investigate bibliometric properties like the number of published papers, the number of active scholars, the average number and the distribution of papers per author, and the average number and the distribution of authors per paper. We study how these properties changed over time in the last 50 years of computer science. 

The second representation of co-authorship that we take advantage of is a coarser network, known in the literature as \textit{collaboration network}, which is obtained from the earlier mentioned affiliation network by an operation of projection on the set of author nodes. A collaboration network is a social network in which the actors are the scholars and the ties represent the collaborations in papers among scholars. In most cases, two authors that have written a paper together do know each other quite well, at least from a scientific perspective, and hence they are tied by a social relationship. This calls for the investigation of typical large-scale properties of social networks, like reachability and average separation distance among scholars, distribution of the number of scholar collaborators, network clustering and network assortativity by number of collaborators. Again, these properties are analysed in a temporal perspective, observing how they varied since the 1960s. Putting together the affiliation and the collaboration network analyses, we get a dynamic picture of how bibliometric and collaboration patterns evolved in the last half-century of computer science.  
          
This is the second part of our investigation of collaboration in computer science using a network science approach. The first part of the study is atemporal: we investigate bibliometric and collaboration properties of the cumulative affiliation and collaboration networks considering all papers published in computer science since 1936 \citep{F11a}. While the first investigation provides a \textit{static} picture of the collaboration in computer science, the present analysis gives a \textit{dynamic} perspective on how collaboration in computer science evolved over time.

\section{Related literature}

Academic collaboration has been extensively studied in \textit{bibliometrics}, the branch of information and library science that quantitatively investigates the process of publication of research achievements; see \citet{CF10} and references therein. Collaboration intensity varies across disciplines: it is fundamental in the sciences, moderate in the social sciences, and mostly negligible in the arts and humanities. Collaborative works are generally valued higher by peer experts and they receive more citations from other papers. 

Collaboration has been also investigated in the context of network science. The notion of academic collaboration network first appeared in a brief note by mathematician \citet{G69}. Newman was the first to experimentally study large-scale collaboration networks with the aid of modern network analysis toolkit. He analysed the structural properties of collaboration networks for biomedicine, physics \citep{N01a,N01b}, and mathematics \citep{N04}, as well as the temporal evolution of collaboration networks in physics and biomedicine \citep{N01c}. \citet{BJNRSV02} studied the evolution in time of collaboration networks in neuroscience and mathematics. The temporal dynamics of mathematics collaboration networks is also investigated by \citet{G02}. \citet{M04} studied the structure and the temporal evolution of a social science collaboration network.

As for studies concerning computer science collaboration networks, \citet{HZLG08} considered publications from 1980 to 2005 extracted from the CiteSeer digital library. The authors studied properties at both the network level and the community level and how they evolve in time. \citet{BBNDFS09} focused on the structure and dynamics of collaboration in research communities within computer science. They isolated 14 computing areas, selected the top tier conferences for each area, and extracted publication data for the chosen conferences from DBLP 2008.  They used network analysis metrics to find differences in the research styles of the areas and how these areas interrelate in terms of author overlap and migration. \citet{MZLA09} made a geographical analysis of collaboration patterns using network analysis. They considered publications from 1954 to 2007 for members of 30 research institutions (8 from Brazil, 16 from North America, and 6 from Europe) and focused on the differences in collaboration habits among these geographical areas. 

Our investigation differs from the mentioned previous studies on computer science for the following reasons: 

\begin{itemize}

\item we use the most complete dataset and build the largest affiliation and collaboration networks ever investigated for computer science; 

\item with the support of the affiliation network representation of collaboration, we study the temporal evolution of bibliometric properties for computer science, like size of the discipline in terms of papers and scholars, author productivity, and collaboration level on papers;

\item with the aid of the collaboration network we study the evolution of meaningful large-scale network properties; in particular the size of biconnected components and the concentration of collaboration using the Gini coefficient have never been examined before for computer science.

\end{itemize}

\section{Methodology}

Data were collected from \textit{The DBLP Computer Science Bibliography} (DBLP, for short) \citep{DBLP}. The DBLP literature reference database was developed within the last 15 years by Dr.\ Michael Ley at Trier University, Germany. DBLP is internationally respected by informatics researchers for the accuracy of its data. As of today, DBLP contains more than 1.6 million entries covering computer and information science. 

Each publication record in DBLP has a key that uniquely identifies the publication and a property that represents the publication type, such as journal article, conference article, book, book chapter, and thesis. Moreover, it contains a semi-structured list of bibliographic attributes describing the publication, like authors, title, and year of publication. This list veries according to the publication type \citep{Ley09}. In particular the publication year attribute was fundamental to make the present time-resolved study.

DBLP is particularly careful with respect to the quality of its data, and is especially sensible to the \textit{name problem}, which includes the cases of a scholar with several names (synonyms) and of several scholars with the same name (homonyms) \citep{RWLWK06}. DBLP uses full names and avoids initials as much as possible. This reduces, but does not eliminate, the name problem. Furthermore, it uses effective heuristics on the collaboration graph to identify possible cases of synonyms or homonyms. For instance, if two lexicographically similar names are assigned to authors that have a distance of two in the collaboration graph, that is, these authors never directly collaborated in a paper but they have a common collaborator, then these names are identified as possible synonyms and they are further manually investigated. Furthermore, if the list of co-authors of an author splits in two or more clusters of highly interconnected authors, but with no collaborations among authors of different clusters, then we might have a case of homonym, and an additional manual check is performed. 

DBLP can be used free of charge. Data can be accessed using a Web interface or through automatic HTTP requests, and the entire dataset can be downloaded in XML format to run experiments on top of it.   

\begin{figure}[t]
\begin{center}
\includegraphics[scale=0.25, angle=0]{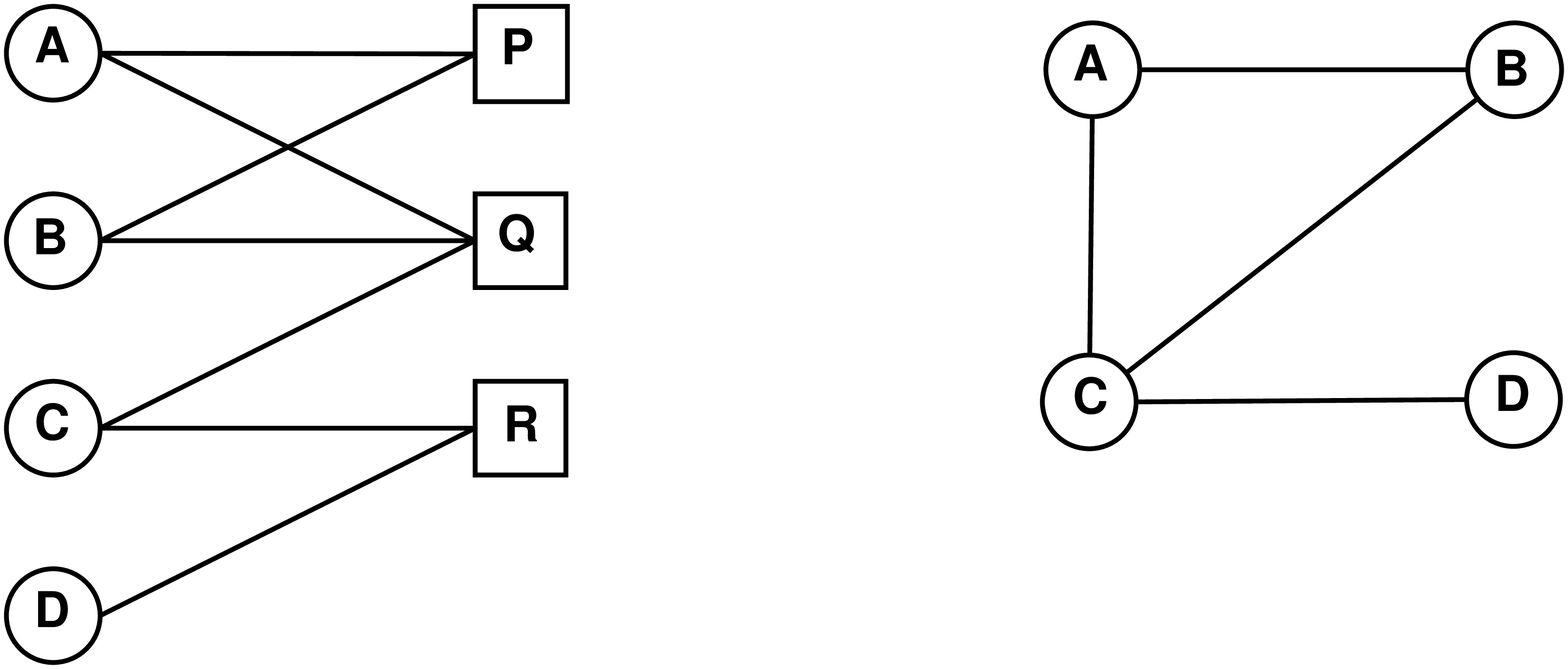}
\caption{A toy example of an author-paper affiliation network (left graph). Authors (circle nodes) match the papers (square nodes) that they wrote.  We also show the corresponding collaboration network (right graph). In this case, two authors are connected if they wrote at least one paper together.}
\label{affiliation}
\end{center}
\end{figure}

We downloaded the XML version of DBLP bibliographic dataset in early 2010 (637.9 MB) and filtered all publications from 1936 to 2008 inclusive.\footnote{We excluded year 2009 since for it the bibliography has not reached the same level of completeness as for previous years.} On top of this database we built the following networks:

\begin{itemize}

\item \textit{Author-paper affiliation network}. This is a bipartite graph with two types of nodes: authors and papers. There is an edge from an author to a paper if the author has written the paper. See an example in Figure~\ref{affiliation}. Affiliation networks are the most complete representations for the study of collaboration \citep{N10}; unfortunately, this bipartite representation for collaboration has received few attention in the literature. For each year $Y$ from 1960 to 2008, we study the \textit{yearly} author-paper affiliation network $AN(Y)$, which contains bibliographic data for all papers published in year $Y$. We excluded years before 1960 since the number of database records for these early years of computer science is not significant.\footnote{For instance, DBLP contains 259 records for year 1959, 23 records for year 1949, and 12 records for year 1936.} 

\item \textit{collaboration network}. A collaboration network is an undirected graph obtained from the projection of the author-paper affiliation network on the author set of nodes. Nodes of the collaboration network represent authors and there is an edge between two authors if they have collaborated in at least one paper.  An example is given in Figure~\ref{affiliation}. Clearly, the collaboration network is a coarser representation with respect to the affiliation network; for instance, if three authors are mutually linked in the collaboration network, then it is not clear, from the analysis of the collaboration network alone, whether they have collaborated in a single paper or in three different ones. Nevertheless, the collaboration network is highly informative since many collaboration patterns can be captured by analysing this form of representation. Furthermore, the collaboration network is the main (mostly unique) representation of collaboration that has been studied in the network science literature. For each year $Y$ from 1960 to 2008, we study the \textit{cumulative} collaboration network $CN(Y)$, which contains collaboration data for all papers published \textit{until} year $Y$. 
\end{itemize}

We saved the collaboration networks in GraphML format (an XML syntax for graphs). We loaded them in the R environment for statistical computing \citep{R} and analysed the structure of the networks using the  R package \textit{igraph} developed by G\'{a}bor Cs\'{a}rdi and Tam\'{a}s Nepusz. On the other hand, we never materialized the (much larger) affiliation networks. Instead, we used XQuery, the standard XML query language, and BaseX \citep{basex}, a light-speed native XML database, to extract the relevant properties from the XML version of the DBLP database.

\section{Affiliation network properties}

In this section we investigate properties of the author-paper affiliation network of computer science. We recall that the computer science author-paper affiliation network is a bipartite graph with two node types representing authors and papers; edges match authors with papers they wrote. For each year from 1960 to 2008, we study the yearly affiliation network build from all papers published in that year. In particular, with the aid of the affiliation network, we investigate the temporal evolution of the following bibliometric properties for computer science: number of published papers, number of active authors, average number of papers per author, and average number of authors per paper. 

\begin{figure}[t]
\begin{center}
\includegraphics[scale=0.40, angle=-90]{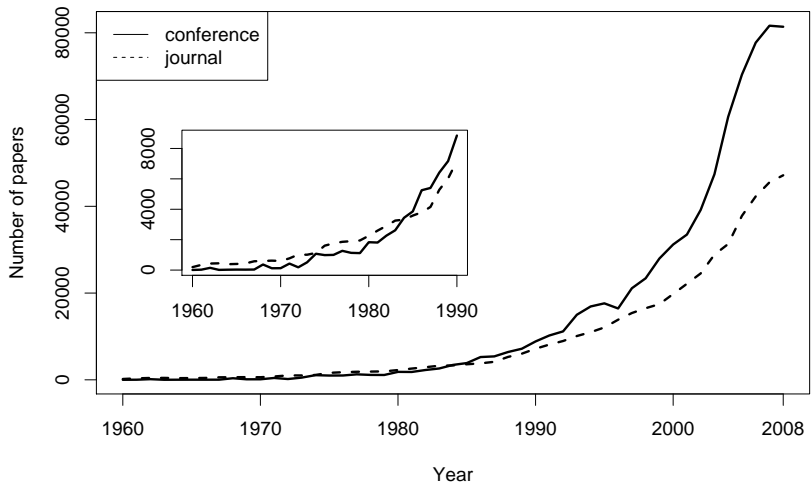}
\caption{The size of the computer science discipline in terms of number of conference papers (solid line) and number of journal papers (dashed line) that are published for each year. The inset refers to the shorter period from 1960 to 1990.}
\label{ppy}
\end{center}
\end{figure}

Figure~\ref{ppy} depicts the size of the computer science discipline in terms of number of papers that are published for each year since 1960. With respect to the author-paper affiliation network, this is the number of nodes of type paper in the network. The computer science field is steadily expanding in terms of number of published papers. However, the proportion of conference and journal papers changed over years. Until 1983, the volume of journal papers dominates that of conference papers. However, since 1984 conferences are the most popular publication venue in computer science, and in the recent years computer science published almost two conference papers every one journal paper \citep{F10-CACM}.

What are the reasons of the growth of the computer science field? We investigated, for each year, the number of active authors, which are authors that published at least one paper in the year, and the author productivity, defined as the average number of papers published by an active author in the year. With respect to the author-paper affiliation network, the number of active authors is the number of nodes of type author in the network, and the author productivity is the average degree of nodes of type author. Indeed, the degree of a node of type author on the author-paper bipartite network is precisely the number of papers published by the author. 

\begin{figure}[t]
\begin{center}
\includegraphics[scale=0.40, angle=-90]{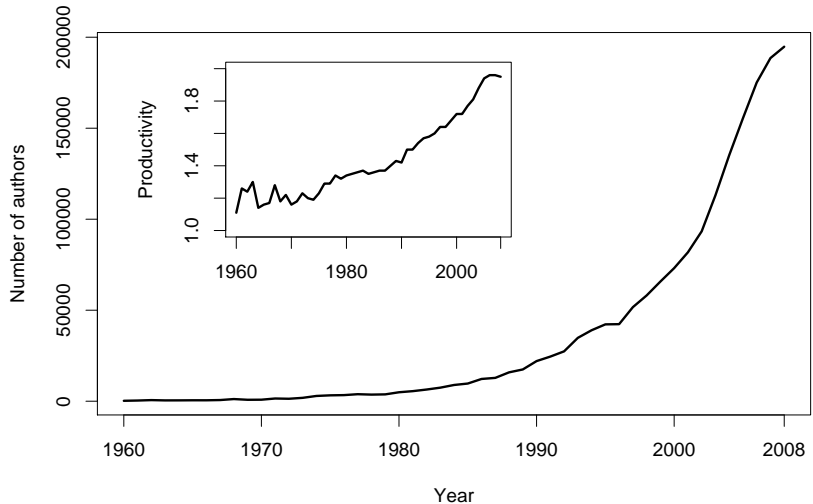}
\caption{The size of the computer science discipline in terms of number of active authors for each year. The inset shows the author productivity in terms of the average number of papers published by an author.}
\label{apy}
\end{center}
\end{figure}

Figure~\ref{apy} shows the temporal evolution of both variables (number of active authors and their productivity). Both variables are growing over time, although the author productivity shows some oscillations during the 1960s. Hence, the expansion over time of computer science in terms of number of papers is justified by an increase in the number of active authors and by a rise of the author productivity. Substantially, the discipline grows since there are more active scholars and these scholars write more papers. 

\begin{figure}[t]
\begin{center}
\includegraphics[scale=0.40, angle=-90]{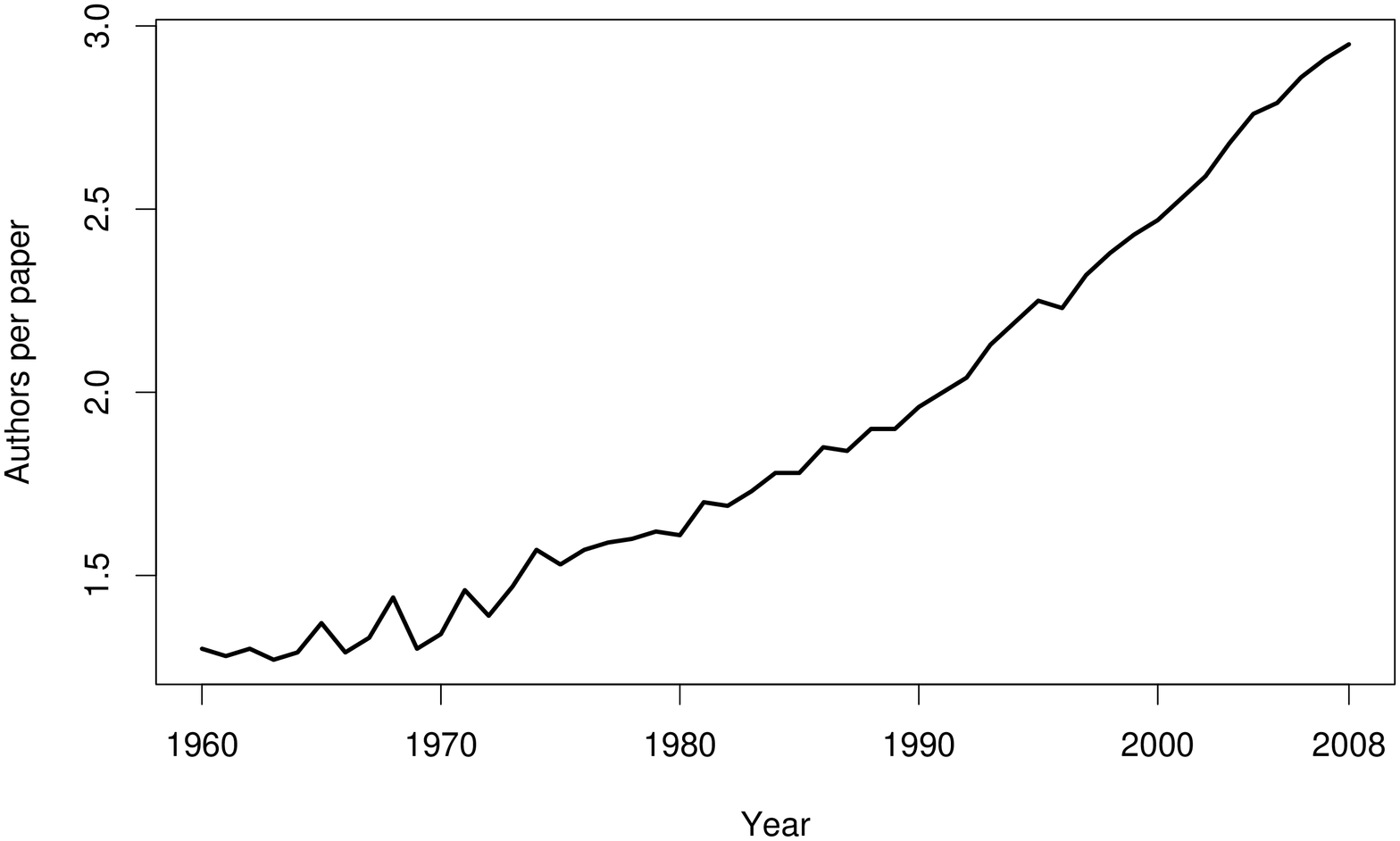}
\caption{The average collaboration level in papers for each year.}
\label{app}
\end{center}
\end{figure}

Finally, we observed, for each year, the average collaboration level in papers, defined as the average number of authors per paper published in the year. With respect to the author-paper affiliation network, the average collaboration level is the average degree of nodes of type paper. Indeed, the degree of a paper node is the number of authors that signed the paper. Figure~\ref{app} shows that collaboration in papers increased over time. The average paper in 2008 has 2.95 authors, while the average one in 1960 had 1.30 authors.\footnote{Recent bibliometric studies have shown that computer science papers jointly written by two authors are generally of better quality (as judged by peer experts) that single-author papers, and the number of citations received from other papers grows with the number of authors of the paper \citep{CF10}.}  

\section{Collaboration network properties}

In this section we study large-scale properties of the collaboration network of computer science. 
Recall that the computer science collaboration network is a graph with nodes representing the computer science scholars and edges representing the collaborations between scholars in research papers. For each year from 1960 to 2008, we study the cumulative collaboration network build from all papers published until that year. Specifically, we investigate the temporal evolution of the following properties of the computer science collaboration network: connectivity of the network, average separation distance among scholars, distribution of the number of scholar collaborators, network clustering and network assortativity by number of collaborators. 

It is reasonable to assume that scientific information flows through paths of the collaboration networks; we expect, indeed, that two authors that collaborated in some paper are willing to exchange scientific information with a higher probability than two scholars that never met. A reasonable question is: is there a collaboration path between any two scholars of the network? If not, what is the largest component of the graph that has the desirable property that any two scholars of the component are connected by at least one path?

A \textit{connected component} of an undirected graph is a maximal subset of nodes such that any node in the set is reachable from any other node in the set by traversing a path of intermediate nodes. A connected component of a collaboration graph is hence a maximal set of authors that are mutually reachable through chains of collaborators. 

A large connected component in the collaboration graph, of the order of the number of scholars, is a desirable property for a discipline that signals its maturity: theories and experimental results can reach, via collaboration chains, the great majority of the scholars working in the field, and thus scholars are scientifically well-informed and can incrementally build new theories and discover new results on top of established knowledge.  Of course, collaboration represents only one way to spread scientific information; the processes of journal publishing and conference attendance make also notable contributions in this direction.

A useful generalization of the concept of connected component in the $k$-connected component (or simply $k$-component). A $k$-component is a maximal set of vertices such that each is reachable from each of the others by at least $k$ node-independent paths. Two paths are said node-independent if they do not share any intermediate node. For the common case $k=2$, $k$-components are called \textit{bicomponents}. An interesting property of $k$-components is that they remain connected (in the usual sense) as soon as less than $k$ nodes (and the incident edges) are removed. For instance, a bicomponent is still connected if we remove one of its nodes. For this reason, the idea of $k$-component is associated with the idea of network robustness. Hence, a bicomponent in a collaboration graph is more robust than an ordinary connected component, since the removal on any scholar with all their collaborators does not destroy the possibility of going from any scholar to any other one via a collaboration path. 

\begin{figure}[t]
\begin{center}
\includegraphics[scale=0.40, angle=-90]{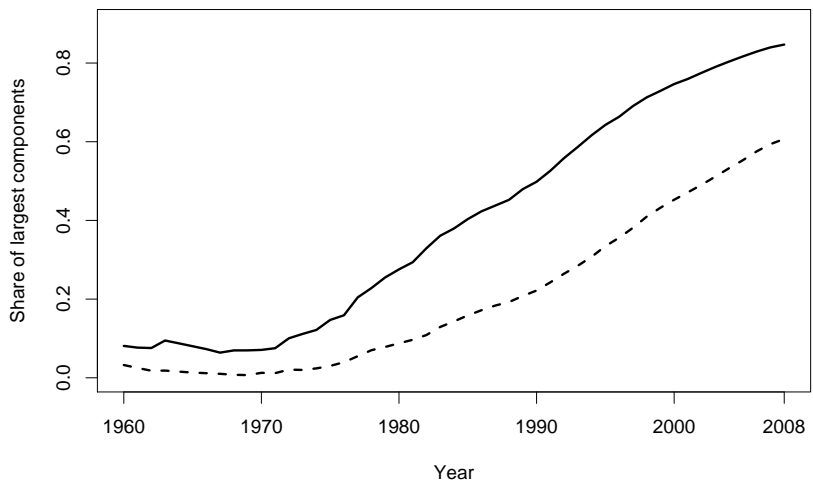}
\caption{The temporal evolution of the share of the largest connected component (solid line) and of the largest biconnected component (dashed line) of the collaboration network.}
\label{cc}
\end{center}
\end{figure}

Figure~\ref{cc} plots the temporal evolution of the share of the largest connected component and of the largest biconnected component of the collaboration network. The fraction of the collaboration network that is taken by the largest connected component grows steadily since 1970. This fraction is below half of the network until 1990, and it covers a share of 80\% in 2004. The figure for the last year, 2008, is 85\%. Notice that the slope of the curve decreases in the last few years, that is the increase of the share of the largest connected component in a given year with respect to the previous year is less noticeable in the recent years. This pattern can be interpreted as a convergence of the collaboration network toward a steady state, where the largest part of the network, although not the entire graph, is connected by paths of collaborations. The line for the largest biconnected component follows a trend similar to the curve for the largest connected component. Not surprisingly, the bicomponent line is below the connected component curve, since any bicomponent is embedded into a connected component. The share of the largest biconnected component continuously grows from the 1970s and reaches half of the collaboration network in 2003, and 61\% of the whole graph in 2008.

The fact that two scholars are connected on a collaboration graph means that there exists at least one path of collaborators between them. However, how long is this path, on average? There exists a substantial difference if the average path connecting two scholars has length, say, six edges, or one hundred links. If we assume that information flows on the network along collaboration ties, then the length of the average path is fundamental to understand how fast the information spreads on the network. 

We may assume that information preferentially flows along \textit{geodesics}, which are shortest paths in terms of number of edges on a graph. The \textit{geodesic distance}, also known as \textit{degree of separation}, between two nodes is defined as the number of edges of any geodesic connecting the nodes. The average geodesic distance is the mean geodesic distance among all pairs of connected nodes of a graph.  The largest geodesic distance in the graph is called the \textit{diameter} of the graph. It tells us how far are two connected nodes in the worst case.

\begin{figure}[t]
\begin{center}
\includegraphics[scale=0.40, angle=-90]{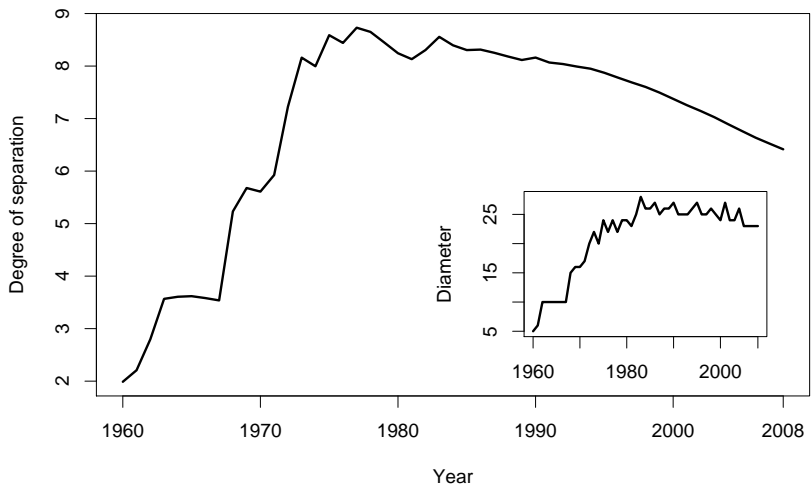}
\caption{The temporal evolution of the average degree of separation between scholars (main plot) and of the maximum degree of separation (inset plot) in the collaboration network.}
\label{geodesic}
\end{center}
\end{figure}

Figure~\ref{geodesic} shows the temporal evolution of the average degree of separation between scholars (main plot) and of the maximum degree of separation (inset plot) in the collaboration network. We can neatly distinguish two phases. The first phase goes from 1960 to 1983. It is characterised by the alternation of periods in which the geodesic distance expands followed by shorter periods in which it shrinks, drawing a curve with a series of ridges. We will see next in this section that the collaboration density among computer scientists, measured as the mean number of collaborators per scholar, continuously increases since 1960. In this initial phase from 1960 to 1983, the collaboration graph is split up in many relatively small connected components, one of them, the largest, collect an increasing share of nodes, from 8\% in 1960 to 36\% in 1983. As the collaboration density increases, it may happen that some of these components glue together to form bigger components, with the effect of raising the average geodesic distance between nodes. These expansion periods are followed by shorter periods in which the collaboration density increases but no significant merging of components occurs, and hence the average geodesic distance between nodes remains stable or more frequently it declines. Since the expansion periods are longer than the contraction ones, in this initial phase the overall average geodesic distance increases, with a maximum of 8.73 reached in year 1977. Also, in this phase, the largest geodesic distance (the diameter) increases, reaching its peak of 28 in year 1983. 

The first phase from 1960 to 1983 is followed by a second phase, from 1983 to 2008, in which the average geodesic distance almost continuously decreases, drawing a valley that gently slopes, as opposed to the sharp ridges of the previous phase. Moreover, the diameter oscillates in a short range (from 23 to 28). In this second phase, there exists a large connected component that contains an important share of the network; this share contains the majority of the nodes starting from 1991. The pairs of nodes in this giant component dominate the computation of the average geodesic distance. Since the collaboration density inside this component becomes higher and higher, the collaboration distances get lower and lower. In 2008, the average degree of separation between computer science scholars in 6.41, a figure that matches well the legendary six degrees of separation found by the experimental psychologist Stanley Milgram in the 1960s with his today highly celebrated small-world experiment \citep{Mi67}. We conjecture that the computer science collaboration network is slowly converging toward a steady state, where the majority of pairs of scholars are connected by short paths (of six or less edges).      

\begin{figure}[t]
\begin{center}
\includegraphics[scale=0.40, angle=-90]{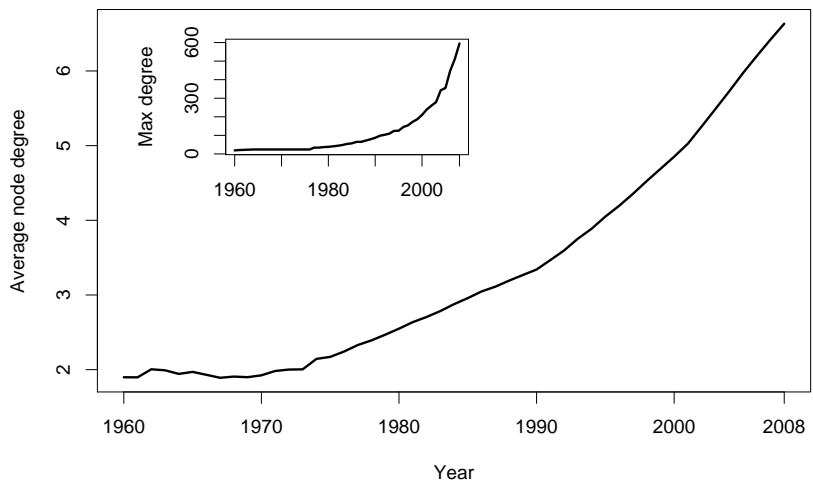}
\caption{The temporal evolution of the average number of collaborators of a scholar (main plot) and of the maximum number of collaborators of a scholar (inset plot).}
\label{degree}
\end{center}
\end{figure}

In the following we investigate how the density and the distribution of the collaboration among computer scientists evolved in time. A measure of the density of collaboration is the average number of collaborators per scholar. This is the average degree of a node in the collaboration graph. 
Figure~\ref{degree} depicts the temporal evolution of the average number of collaborators of a scholar (main plot) and of the maximum number of collaborators of a scholar (inset plot). The mean number of collaborators is stable in the 1960s, with an average of 1.9 co-authors per scholar. From 1970 it steadily increases: the average computer scientist has 2.2 collaborators in the 1970s, 2.9 collaborators in the 1980s, 4 collaborators in the 1990s, and 5.7 collaborators in the 2000s. The mean number of collaborators in the last year 2008 is 6.6, more than three times the number of collaborators a computer scientists had in the 1970s. The maximum number of collaborators of a scholar, although a less significative figure, is also increasing with time, reaching the impressive maximum amount of 595 collaborators for a single scholar in 2008.\footnote{As a comparison, the most prolific and the most collaborative among mathematicians, Paul Erd\H{o}s, wrote more than 1400 papers cooperating with more than 500 co-authors \citep{G97}.}

We have just noticed that the number of collaborations increased over time. However, did the distribution of collaborations among scientists change over time? \textit{Concentration} measures how the character (in our context, the collaborations) is equally distributed among the statistical units (the scholars). The two extreme situations are equidistribution, in which each statistical unit receives the same amount of character (each scholar has the same number of collaborators) and maximum concentration, in which the total amount of the character is attributed to a single statistical unit (there exists a super-star collaborator that co-authored with all other scholars, and each other scholar collaborated only with this collaboration super-star). Since a collaboration is represented with an edge in the collaboration graph, the concentration of collaborations among computer scientists corresponds to the concentration of edges among nodes in the collaboration graph. 

\begin{figure}[t]
\begin{center}
\includegraphics[scale=0.40, angle=-90]{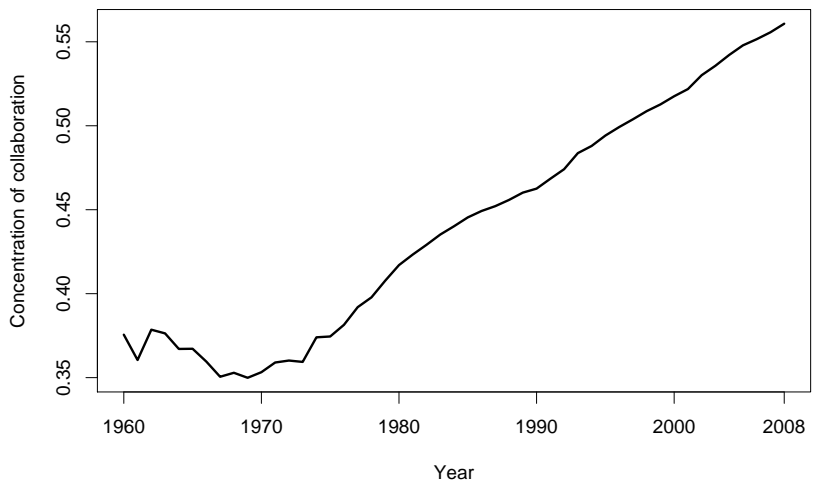}
\caption{The temporal evolution of the collaboration concentration measured with the Gini concentration coefficient.}
\label{gini}
\end{center}
\end{figure}

A numerical indicator of concentration is the \textit{Gini concentration coefficient}, which ranges between 0 and 1 with 0 representing equidistribution and 1 representing maximum concentration. Figure~\ref{gini} shows how this coefficient varied with time in the computer science collaboration network. From 1970 the concentration of collaboration among our peers moved away from the equidistribution toward a more concentrated scenario. The Gini collaboration coefficient for the last surveyed year is 0.56. In this year, the most collaborative 1\% of the scholars harvest 13\% of the collaborations, the 5\% of them collect one-third (33\%) of the collaborations, and the 10\% of them attract almost half (46\%) of the collaborations. Most of the collaborations are achieved by few vital scholars, with the majority of scholars collecting the minority of the collaborations. Hence the distribution of collaborations over scholars is highly skewed, as expected, and the distribution skewness has increased over time. 

Finally, we focus on the topological structure of the computer science collaboration network. To this end we use two apparently unrelated measures: the transitivity coefficient and the assortativity coefficient. The \textit{transitivity coefficient}, also known as \textit{clustering coefficient}, measures the average probability that a collaborator of my collaborator is also my collaborator, and it can be computed as the ratio of the number of loops of length three and the number of paths of length two in the collaboration graph. The coefficient ranges from 0 to 1; a value of 1 implies perfect transitivity, i.e., a network whose components are all cliques.\footnote{A clique is a graph in which each pair of nodes is connected by an edge.} A value of 0 implies no closed path of length two, which happens for various topologies, such as a tree or a square lattice. 

Assortative mixing by degree is the tendency of nodes to connect to other nodes with a similar degree. In our context, we have assortative mixing by degree if scholars collaborate preferentially with other scholars with similar number of collaborators. We have disassortative mixing by degree if collaborative scholars co-author with hermits and vice versa. We have no mixing at all if none of these patterns is clearly visible. A quantitative measure of the magnitude of mixing by degree is the \textit{assortativity coefficient}, which is the Pearson correlation coefficient applied to the degree sequences of nodes connected by an edge. The coefficient ranges between -1 and 1, where negative values indicate disassortative mixing, positive values indicate assortative mixing, and values close to 0 indicate no mixing.

\begin{figure}[t]
\begin{center}
\includegraphics[scale=0.40, angle=-90]{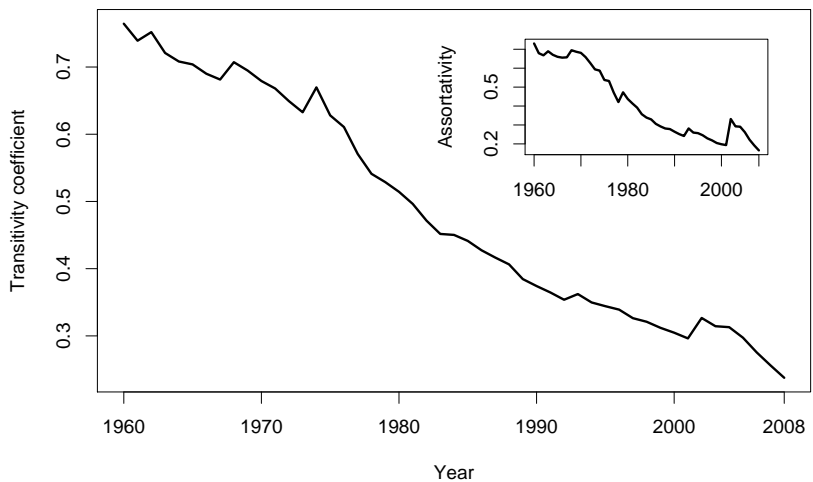}
\caption{The temporal evolution of the transitivity coefficient (main plot) and of the assortativity coefficient (inset plot) of the collaboration network.}
\label{transitivity}
\end{center}
\end{figure}

Figure \ref{transitivity} illustrates the transitivity coefficient (main plot) and of the assortativity coefficient (inset plot) of the collaboration network. Substantially, the transitivity coefficient is decreasing in time, ranging from a value of 0.76 in 1960 to a value of 0.24 in 2008.\footnote{Notice, however, that the transitivity coefficient of 0.24 for the 2008 collaboration network is still a rather high probability, especially when compared with non-social networks or with random networks. As a comparison, the transitivity coefficient for a network with the same degree distribution of the 2008 collaboration network but otherwise random is $0.00014$.} In 1960 it holds that 3 times over 4 two scholars sharing a common collaborator are themselves co-authors. After 50 years of computer science, the same phenomenon happens only 1 time over 4.

As to assortativity, in the early years, the collaboration network is strongly assortative by degree, with an assortativity coefficient of 0.73. The collaboration network remains assortative by degree in the following years, but the strength of assortativity decreases, and reaches its minimum in 2008 with an assortativity coefficient of 0.17.\footnote{The assortativity coefficient of 0.17 is still a large one if compared with non-social networks, which typically show null or negative assortativity coefficients.}

\begin{figure}[t]
\begin{center}
\includegraphics[scale=0.4, angle=-90]{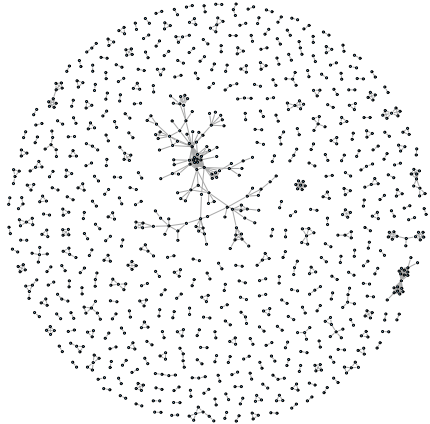}
\caption{The computer science collaboration networks in 1964.}
\label{1964}
\end{center}
\end{figure}

\begin{figure}[t]
\begin{center}
\includegraphics[scale=0.4, angle=-90]{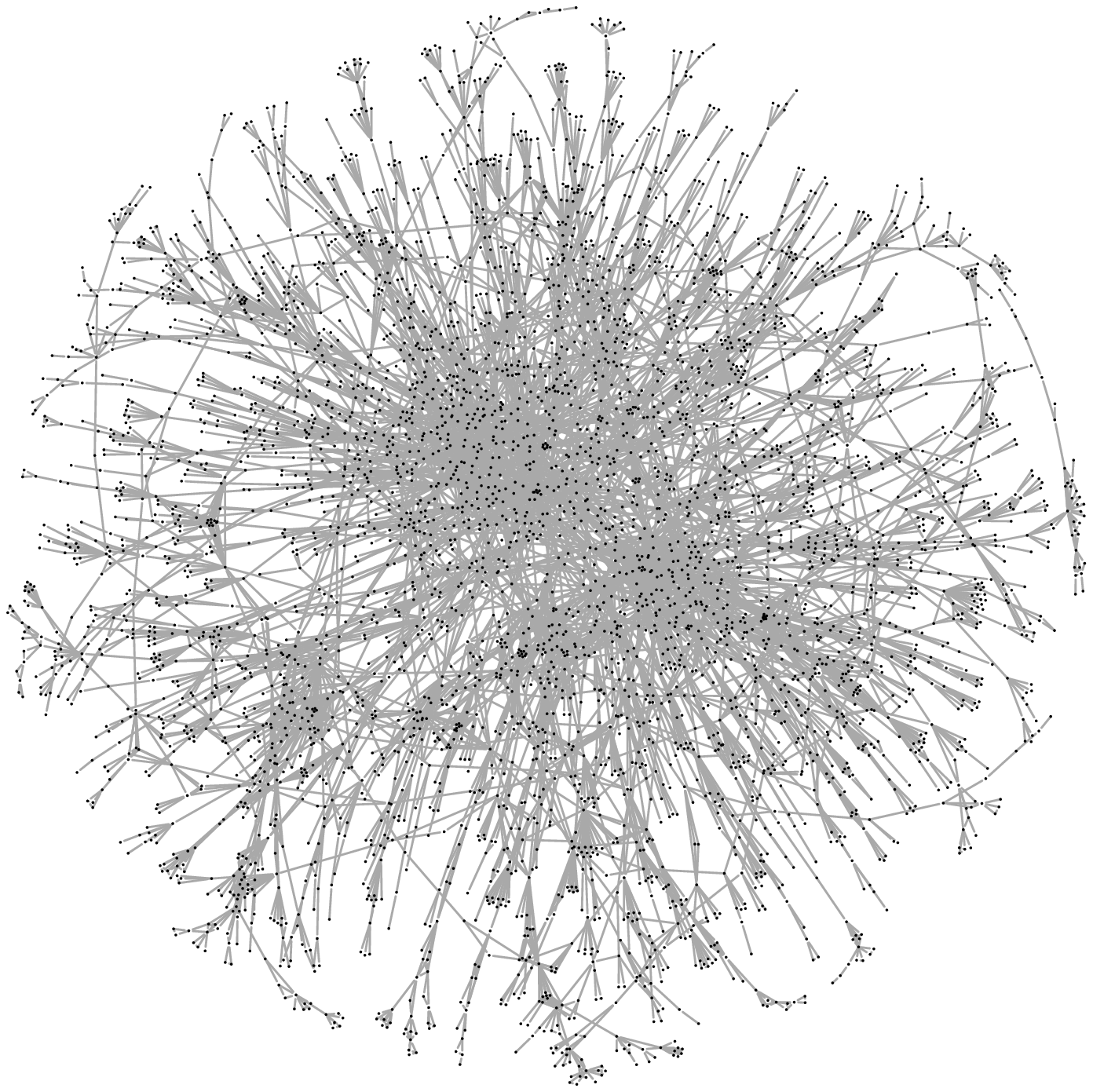}
\caption{The computer science collaboration networks in 1980 (only the largest component is shown).}
\label{1980}
\end{center}
\end{figure}

Interestingly, the assortativity coefficient and the transitivity coefficient follow very similar trends and seem to be highly correlated. Indeed, the Pearson correlation coefficient between the two variables is 0.98, which means almost perfect positive correlation.\footnote{To the knowledge of the writer, this is the first time that such strong association between transitivity and degree assortativity is observed in the evolution of real networks.} The synchronous behaviour of the transitivity and of the assortativity coefficients reveals an interesting change in the topology of the collaboration network over time. In the early times of computer science, the collaboration network had a clear \textit{core-periphery structure}, with a core of interconnected high-degree nodes surrounded by a less dense periphery of nodes with lower degree. In particular, the periphery of the network is composed of small clusters of nodes, which are internally highly connected, but disconnected from the other small groups and from the core of the network. See Figure~\ref{1964}, depicting the collaboration network as it was in 1964, for an example. This kind of network has both a high clustering coefficient and a high assortativity coefficient. Indeed, the core of the network is highly clustered, and so are also the peripheral small clusters, and hence the clustering coefficient is high. Moreover, the central high-degree nodes of the core are interconnected, and the peripheral low-degree nodes match with other similar nodes, making the network highly assortative by degree.

The computer science collaboration network progressively lost this peculiar core-periphery structure over time, or at least this structure is not so clear in the years of modern computer science. The modern structure of the collaboration network is dominated by the largest connected component of the network, which covers the majority of the nodes. The disconnected periphery is still composed of small clusters of nodes, but, unlike in the early years, this part of the graph contains an insignificant share of nodes. The nodes in the largest component are not as highly linked as were the nodes in the core component of early networks. Moreover, highly collaborative scholars in the modern collaboration network are more willing to collaborate with those with few collaborators (as in the typical relationship supervisor-student). Figure~\ref{1980} shows the largest component for the 1980 collaboration network. Notice that, to a certain degree, the core-periphery structure is still noticeable, although it is less clear than in the core component of Figure~\ref{1964}. 

\section{Conclusions}

We have analysed how collaboration in computer science evolved in the last half-century using a network science approach. Computer science is expanding since 1960 both in terms of number of published papers and number of active authors. Moreover, computer scientists became more productive and collaborative. Productivity and collaboration are mutually reinforcing, since more collaborative authors generally write more papers and more productive scholars typically attract more collaborations. Collaboration changed not only in intensity, but also in concentration: the gap between the richest and the poorest scholars, in terms of collaborations, is increasing. The temporal evolution of the collaboration network is approaching a steady state, where most of the scholars belong to a giant connected component and are separated by few collaboration links. Finally, the collaboration network moved from a clear core-perithery topology to a more balanced one, with lower transitivity and lower homophily by number of collaborators. In particular, the lower transitivity with respect to the early times of computer science might indicate a higher propensity for collaborations among different research institutes, among different countries, and among different sub-fields of computer science. Triads involving such heterogeneous collaborations, indeed, are less likely to be closed in a collaboration triangle. Recent bibliometric studies have shown that papers involving collaborations over different institutes, countries, or research fields generally enjoy a higher visibility and attract more citations than papers with homogeneous authors, making these types of collaborations more appealing \citep{CF10}. 

\bibliographystyle{elsarticle-harv}
\bibliography{bibliometrics}

\end{document}